\documentclass{elsart}
\usepackage{graphicx, amssymb}
\journal{Nuclear Instruments and Methods A}
\begin{document}
\begin{frontmatter}

\title{A two dimensional position sensitive
gas chamber with scanned charge transfer readout}

\author[facul]{F. G\'omez \corauthref{cor}},
\corauth[cor]{Corresponding author.}
\ead{faustgr@usc.es}
\author[facul]{A. Iglesias},
\author[hospi]{R. Lobato},
\author[hospi]{J. Mosquera},
\author[facul]{J. Pardo},
\author[facul]{J. Pena},
\author[facul]{A. Pazos},
\author[hospi]{M. Pombar},
\author[facul]{A. Rodr\'{\i}guez}

\address[facul]{Universidade de Santiago,
Departamento de F\'{\i}sica de Part\'{\i}culas}
\address[hospi]{Hospital Cl\'{\i}nico Universitario de Santiago}
\thanks{This work was supported by project PGIDT01INN20601PR from Xunta de Galicia}

\begin{abstract}
We have constructed and tested a 2d position sensitive
parallel-plate gas ionization chamber with scanned charge transfer readout.
The scan readout method described here is based on the development of a new position dependent charge transfer technique.
It has been implemented by using gate strips perpendicularly oriented to the collector strips.
This solution reduces considerably the 
number of electronic readout channels needed to cover large detector areas. The use of a 
25$\mu$m thick kapton etched circuit  allows
high charge transfer efficiency with
a low gating voltage,consequently needing a very simple commutating circuit.
The present prototype covers 8$\times$8 cm$^2$ 
with a pixel size of $1.27 \times 1.27$ mm$^2$.
Depending on the intended use and
beam characteristics a smaller effective pixel is feasible and larger active areas are possible.
This detector can be used for X--ray 
or other continuous beam intensity profile monitoring. 
\end{abstract}


\end{frontmatter}

\section{Introduction}

To monitor high rate beam fluency gas ionization chambers based devices \cite{ahmed} can be used.
In order to obtain a 2d profile of radiation dose or fluency it is necessary to have a highly segmented
anode \cite{listin} \cite{besch}. 
This represents a major problem when the size of the chamber is large compared
with the anode segmentation pitch. In the present work we describe a new and simple  solution 
developed to give real two dimensional readout on a gas ionization chamber. 
The method consists on scanning electronically the detector active area
in the perpendicular direction to the readout strips orientation (strips instrumented
with the measuring electro-meters). Gated charge transfer 
is a well known method which has been used in different 
radiation detectors like, for example,  
gated gas-filled Time Projection Chambers \cite{tpc} in High Energy Physics.

\begin{figure}
\begin{center}
\includegraphics*[width=10cm]{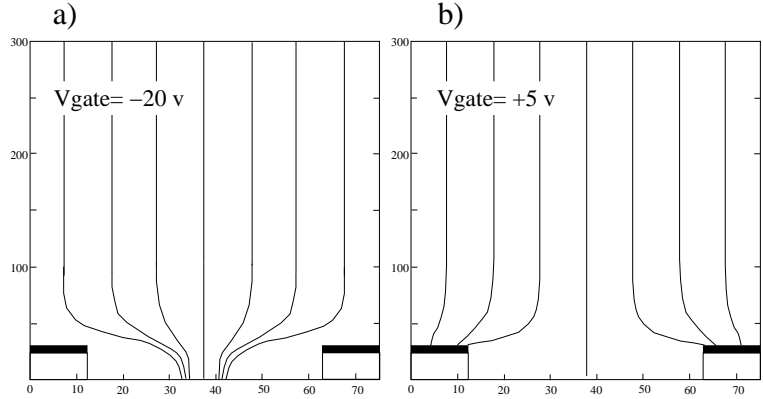}
\end{center}
\caption{Working principle of the charge transfer readout.
Electric field lines map for the configuration of {\em "open gate"} {\bf a)} and
{\em "closed gate"} {\bf b)}. The drift field value used in the simulation is 1.3$\times$$10^5$ V/m. 
Dimensions are given in $\mu$m.}
\label{puerta}
\end{figure}

The transfer of electrons drifting from the detector volume to the collector electrode strips can be blocked
by means of a local electric field that dominates the drift field in the proximity 
of the gate electrode strips (see simulation in Figure \ref{puerta}). 
In this way we provide a single stage control of the charge transfer from the
gas drift region to the collector electrodes. 
In this work we profit from the great development of Micro--Pattern Gas Detectors (MPGD) 
\cite{sauli} in the last years, specially the advances on kapton etching techniques.
In our case we integrated a gating grid using a two metal layer circuit based in a 
kapton foil glued on a FR4 epoxy substrate to provide mechanical rigidity.
The use of a 25$\mu$m thick kapton foil to produce a two level readout circuit 
allows to use a small gating voltage.

\section{Prototype description}
To design and optimize the geometry of the gate and collector electrodes we 
simulated the electric field map of a transfer gap cell through MAXWELL3D\footnote{Ansoft
Corporate, Pittsburgh USA} and MATLAB\footnote{The MathWorks, Inc. 
3 Apple Hill Drive, Natick MA(USA) 01760-2098} 
programs. For the final design it was chosen a 1:2:1 ratio corresponding to 
the transfer gap height:gate--spacing:gate--width. With this geometry we expected to have a low
electron transfer transparency (well bellow 1\%) using a relatively low positive (+5V) {\em "closed gate"} 
voltage\footnote{In the present work collector electrodes are always grounded.} and moderate drift fields ($\sim$ $10^5$ V/m).
A wider gate strip spacing would increment the electron transparency  
for a given negative {\em "open gate"} voltage, but  also higher positive voltages would be needed to block
efficiently the electron drift.
On the other hand the local electric field present around the gate strip either in open or closed mode should not affect
significantly the electric field in the drift region. 
This is another important reason to use a small transfer gap between the gate and collector strips.

\begin{figure}
\begin{center}
\includegraphics*[width=10cm]{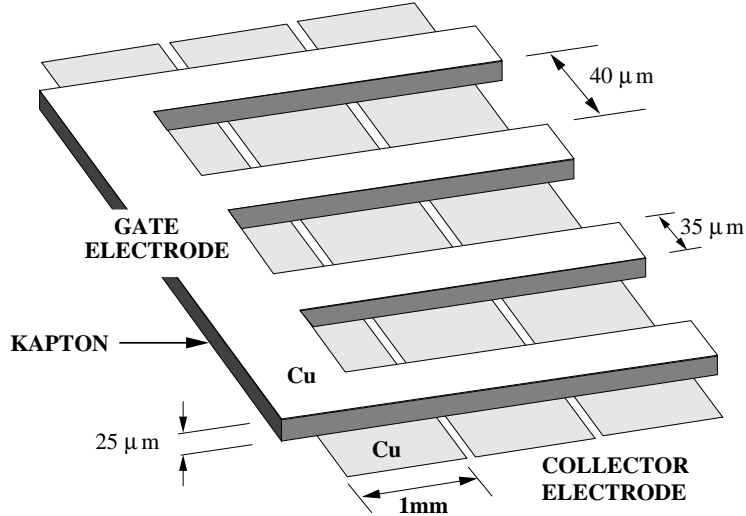}
\end{center}
\caption{Scheme of the gate and collector electrode layout. In the gate 
electrode 17 strips were joined in a group to provide a 1.27mm pitch both in
the X and Y directions. Dimensions in the drawing are not to scale.}
\label{esquema}
\end{figure}
\begin{figure}
\begin{center}
\includegraphics*[width=10cm]{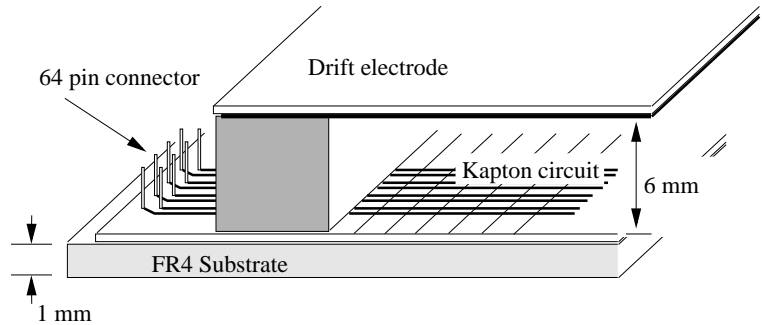}
\end{center}
\caption{Gas ionization chamber layout used for this prototype} 
\label{layout}
\end{figure}

The readout circuit layout of this prototype is described in Figure \ref{esquema}.  
We used a 25$\mu$m thick kapton foil metal coated on both sides with 5$\mu$m copper.
After etching of the copper, we obtained 35$\mu$m wide and 1 mm wide strips on the gate and collector plane respectively. 
The foil was glued on a 1mm FR4--epoxy substrate and
subsequently kapton in the gap between gate strips was removed obtaining the Figure 
\ref{esquema} micro-pattern. In order to provide a simple connection procedure 
a  64 pin 1/10 inch pitch standard connector was included
in the board edge design for both gate and collector strips. This choice implies
a 1.27 mm pitch in the detector active area. The collector strips were thus made
1mm wide with 1.27 pitch while 17 gate strips were joined in a group in the 
detector edge to achieve the previous effective pitch.
The drift gas gap was built using a G10 6mm height frame glued on the readout board,
and this volume was closed with a drift electrode made of 200$\mu$m thick G10 copper clad on one side.

\begin{figure}
\begin{center}
\includegraphics*[width=10cm]{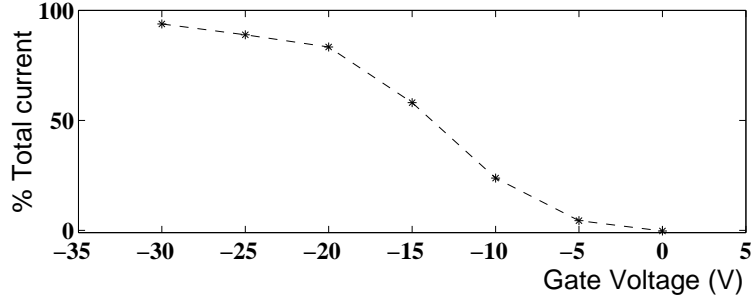}
\end{center}
\caption{Fraction of the total ionization current arriving to the
collector electrode as a function of the gate voltage. This measurements
were done using pure Ar with an electric drift field of $2 \times 10^4$V/m.}
\label{conmutacion}
\end{figure}

\section{Charge transfer}

The experimental transparency for electrons obtained as a function of the gate voltage is shown 
in Figure \ref{conmutacion}. This measurements were done using pure Ar at atmospheric 
pressure, and corresponds to the ratio of the ionization current  under X--ray irradiation 
measured at the collector and at the drift electrode.
The plot shows that very small transparency values ($\sim 10^{-3}$) can be obtained for low positive voltages at {\em "closed mode"}. 
The transparency increases when applying negative voltages at the gating electrodes, showing that there is a transition region 
centered on -15V, and can reach values very close to 1 ({\em "open mode"}) when the gate voltage is below -20V. 
If the drift field is increased, the gate voltage has also to be increased proportionally to maintain the
same transparency value. The fraction of current produced by electrons arriving to the gate electrode is
the complementary value of the transparency calculated for the collector.
We have chosen -17V as {\em open gate} voltage as this value can be commutated using standard CMOS analogue switches. 
If we consider the width of the gate strips $w$ (35 $\mu$m in Figure \ref{esquema}) 
and the corresponding space between gate strips $c$ (40 $\mu$m in Figure \ref{esquema}),
the expected electron transparency $t$ is equal to the optical limit defined by

$$t=\frac{c}{c+w}= 0.53\;\;\;\;\;  {\rm at} \;\;\;\;\;    {V_g}={g}\frac{V_d}{h}= - 0.5V$$ 

being $V_g$ and 
$V_d$ the gate and drift voltages respectively, 
$g$ the height of the gate strips over the collector electrodes and $h$ the gas gap below the 
drift electrode. The measured maximum transparency significantly deviates from this expectation due to the trapezoidal
etching of the kapton layer that partially covers the collector electrodes in the space between gate 
strips.

\section{Two dimensional readout}

To demonstrate the working principle of our position sensitive charge transfer scan readout method, a small area
of the prototype was instrumented with charge integration electronics. To drive the detector, two separate
electronic boards were used, one with the commutation circuit to sequentially commutate 
from closed to open gate voltage on the gate strips and another board with the necessary electronic readout channels to integrate the 
charge collected by each individual collector strip (integrators board). 
To integrate the collector current we used a precision switched IVC102 \cite{ivc} integrating amplifier 
(with an internal feedback capacitor set to 10 pF).

\begin{figure}
\begin{center}
\includegraphics*[width=14cm]{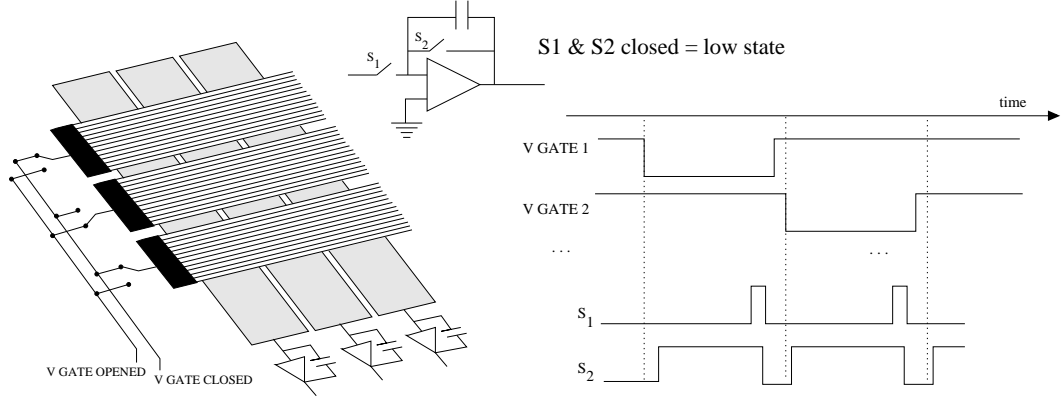}
\end{center}
\caption{Scheme of the 2d scan readout. By default all the gate electrodes are connected to a bias voltage (positive, closed gate voltage) 
giving low electron transparency.
During the readout cycle each individual gate electrode strip is sequentially set to a voltage value (negative, open gate voltage) 
that modifies the electric field around the gate strip.  This allows the transfer to the collector plane of a large fraction of the electrons present at 
the volume defined by the drift electrode a the selected gate strip. 
A typical readout time sequence is shown. }
\label{esquema2}
\end{figure}

To sequentially change the voltage value applied to each gate strip a shift register was used, built by means of daisy chained
D--type flip--flops (HEF40174B) whose outputs were connected to double analogue switches
(AD7592DI). These analogue switches commute from  closed gate voltage value (that keeps the 
electron transparency at lowest values), to a negative open gate voltage value giving a high electron transparency for the selected gate strip
(typically $>$70\% in our setup). A scheme of the readout circuit is shown on Figure \ref{esquema2}.
For the synchronous control of the two electronic boards we used a personal computer PCI 
embedded card\footnote{National Instruments PCI-6023E DAQ card.}. 

The first prototype was instrumented with a total number of 96 effective pixels (on an active area of 15.2 $\times$ 10.2 mm$^2$), 
corresponding to 12 gate channels and 8 collector channels. 
The total capacitance at the detector between gate and collector plane is 7nF, meaning a 2pF capacity per pixel.   
This gate--collector capacity gives a 2$\mu$s transient time during commutation, considering the 3k$\Omega$ impedance
of the IVC102 (with switches S1+S2 closed).   
This dead time is small compared with the typical integration times used (of the order of milliseconds).
The total time required to obtain an image is equal to the integration time needed to integrate the charge transfered by each gate strip
times the number of gate strips.

\begin{figure}
\begin{center}
\includegraphics*[width=15cm]{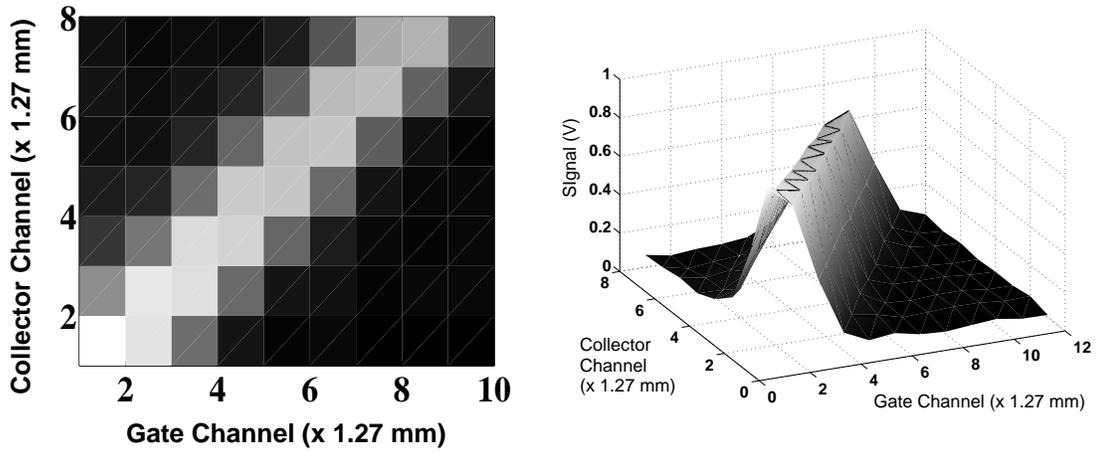}
\end{center}
\caption{ 2d and 3d dose distribution of a 1.5 mm slit collimator inclined 45$^0$ irradiated with soft 
X--rays.}
\label{ranura}
\end{figure}

\begin{figure}
\begin{center}
\includegraphics*[width=5cm]{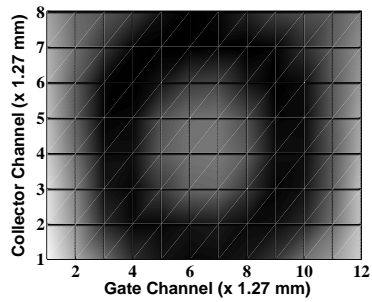}
\end{center}
\caption{ X--ray image of a screw nut with an internal hole of 5mm diameter. 
Pixel values were linearly interpolated to generate this image.}
\label{anillo}
\end{figure}

Figure \ref{ranura} and \ref{anillo} show two X--ray images obtained with this prototype using a Chromium X--ray tube: a 1.5 mm slit between
two 5mm thick aluminum plates and a 5mm screw nut were illuminated. 
The $closed$ and $open$ gate voltage values used were +5V and -17V respectively, and the drift field applied in the image was 1.7$\times$10$^4$ V/m in pure Ar.
To obtain this images a scan readout cycle time of less than 10 seconds was required, with a collector current value of 7 pA in the pixels with maximum signal.

The charge transparency value when the gate electrode is set to the 
{\em closed } state should be ideally zero. But the real value differs from 
zero by a small amount ($i.e.$ 0.1\%). This causes a leakage
current, of a value proportional to the detector area, that can seriously distort the image. Nevertheless,
if the leakage current per effective pixel is small we can correct this effect by what we have 
called {\em differential readout}. The correction term is measured by integrating the 
detector current during the same time interval used for the standard 
readout with all the gate electrodes {\em closed}. 
This value will then be subtracted as a pedestal from the values obtained with the gate
electrodes on {\em opened} state. In this way the image will 
not be dramatically affected by the non zero transparency value at the {\em closed } gate
state. 

\section{Conclusions}

We have proved the working principle of a simple and reliable readout method for 2d position sensitive 
gas ionization detectors. 
This two dimensional charge transfer readout solution allows to cover large detector areas giving a high number of effective pixels 
and minimizing the number of readout electronic channels. 
By using a kapton--insulated two layer readout circuit, charge transfer does not affect the drift field and 
the control of the charge transfer process can be done with very low voltage values. The detector can be 
used not only with gases but with other photo-conductive media like non--polar 
liquids ($i.e.$ isooctane or tetrametilsilane). For low beam intensity or fast
readout cycle applications, an intermediate gas avalanche device, like a Gas Electron 
Multiplier \cite{gem}, can be added to improve its sensitivity. 
Considering moderate gas gains around 100, an image could
be readout with a cycle period 100 times faster than
an equivalent ionization chamber, leading to a fast beam imaging device.

\section{Acknowledgments}

We are grateful to Manuel Sanchez from CERN EST/DEM group for his permanent 
technical support and collaboration.


\begin{thebibliography}{3}

\bibitem{ahmed} S.N. Ahmed, H.-J. Besch, A.H. Walenta, N. Pavel 
and W. Schenk, Nucl. Instr. and Meth. A 449(2000) 248

\bibitem{listin} V.K. Myalistin, H.-J. Besch, H.W. Schenk 
and A.H. Walenta, Nucl. Instr. and Meth. A 323(1992) 97

\bibitem{besch} H.J. Besch, E.J. Bode, R.H. Menk, H.W. Schenk, U. Tafelmeier,
A.H. Walenta and H.Z. Xu, Nucl. Instr. and Meth. A 310(1991) 446


\bibitem{tpc} The ALEPH collaboration, Nucl. Instr. and Meth. A 294(1990) 121

\bibitem{sauli} F. Sauli, A. Sharma, Ann. Rev. Nucl. Part. Sci. 49 (1999) 341

\bibitem{ivc} Precision switched integrator transimpedance amplifier. Data sheet, 
Burr--Brown, USA 2000.

\bibitem{gem} S. Bachmann, A. Bresan, S. Kappler, B. Ketzer, M. Deutel, L. Ropelewski,
F. Sauli, E. Schulte Nucl. Instr. and Meth. A 471(2001) 115

\end{thebibliography}
\end{document}